\documentstyle[12pt]{article}

\begin{document}
\setcounter{page}{1}

\pagestyle{plain}
\vspace{1cm}
\begin{center}
\Large{\bf Wave Packets Propagation in Quantum Gravity}\\
\small
\vspace{1cm} {\bf Kourosh Nozari$^{1,2}$}\quad and \quad {\bf S. Hamid Mehdipour$^{1}$}\\
\vspace{0.5cm} {\it $^{1}$Department of Physics,
Faculty of Basic Sciences,\\
University of Mazandaran,\\
P. O. Box 47416-1467, Babolsar, IRAN\\ $^{2}$Research Institute for
Astronomy and Astrophysics of
Maragha, \\P. O. Box 55134-441, Maragha, IRAN\\

e-mail: knozari@umz.ac.ir}
\end{center}
\vspace{1.5cm}

\begin{abstract}
Wave packet broadening in usual quantum mechanics is a consequence
of dispersion behavior of the medium which the wave propagates in
it. In this paper, we consider the problem of wave packet broadening
in the framework of Generalized Uncertainty Principle(GUP) of
quantum gravity. New dispersion relations are derived in the context
of GUP and it has been shown that there exists a gravitational
induced dispersion which leads to more broadening of the wave
packets. As a result of these dispersion relations, a generalized
Klein-Gordon equation is obtained and its interpretation is given.
\end{abstract}
\newpage
\section{Introduction}
In probabilistic interpretation of usual quantum mechanics, wave
packet broadening is described as reduction of the probability of
finding a particle in a given volume at a given time. This reduction
of probability or broadening is related to dispersive nature of the
medium which wave propagates in it[1]. Consider a more realistic
situation which incorporates gravity with quantum theory. This
situation has very important candidates such as early Universe and
massive black holes interior. In this circumstances the usual
uncertainty principle of Heisenberg should be generalized to
incorporate gravitational uncertainty from very beginning . This
generalized uncertainty principle (or more reasonable terminology of
Gravitational Uncertainty Principle) leads to many interesting
results in Planck scale physics[2-15]. In this paper we consider the
problem of wave packet propagation in GUP. We will show that there
is an additional broadening due to gravitational effects and this is
a consequence of generalized dispersion relation in GUP. Such a
modified dispersion relation leads to a modified form of
Klein-Gordon equation. Generalized form of Klein-Gordon equation
suggests that one can define a generalized momentum operator and
this generalization of momentum operator can be interpreted as
generalized De Broglie principle which is the foundation of quantum
theory. The other possible interpretation of this generalized
momentum operator is related to the fact that $\hbar$ may be a
varying "constant" with wave vector. Our calculations show that in
Planck scale, the group velocity of the wave packet can be greater
than light velocity and this is not surprising since recently
variation of fundamental constants of the nature is accepted, at
least for fine structure constant which contains both light velocity
and $\hbar$[16-20].\\
The structure of the paper is as follows: in section 2 first we give
a short outline to wave packet propagation in usual quantum
mechanics and then our calculation for wave propagation in GUP is
given. Section 3 gives a generalized Klein-Gordon equation and a
generalized Momentum operator. Summary and Conclusions are given in
section 4.
\section{Wave Packet Propagation}
\subsection{ Wave Packet Propagation in Ordinary Quantum Mechanics}
Consider the following plane wave profile,
\begin{equation}
\label{math:1.1} f(x,t)\propto e^{\, ikx-i\omega t}.
\end{equation}
Since  $\omega= 2\pi\nu$,  $k=\frac{2\pi}{\lambda}$  and
$\nu=\frac{c}{\lambda}$,  this equation can be written as
$f(x,t)\propto e^{\, ik(x-ct)}$. Now the superposition of these
plane waves with amplitude $g(k)$ can be written as,
\begin{equation}
\label{math:1.2}f(x,t)=\int_{-\infty}^\infty dk\, g(k)\, e^{\,
ik(x-ct)}=f(x-ct)
\end{equation}
where $g(k)$ can have Gaussian profile. This wave packet is
localized at $x-ct=0$. In the absence dispersion properties for the
medium, wave packet will not suffers any broadening with time. In
this case the relation $\omega=kc$ holds. In general the medium has
dispersion properties and therefore $\omega$ becomes a function of
wave number, $\omega=\omega(k)$. In this situation equation (2)
becomes,
\begin{equation}
\label{math:1.3}f(x,t)=\int dk\, g(k)\, e^{\, ikx-i\omega(k)t}.
\end{equation}
Suppose that $g(k)=e^{-\alpha(k-k_{0})^{2}}$. With expansion of
$\omega(k)$ around $k=k_{0}$, one find
\begin{equation}
\label{math:1.4}\omega(k)\approx
\omega(k_0)+(k-k_0)\bigg({\frac{d\omega}{dk}}\bigg)_{k_0}+\frac{1}{2}(k-k_0)^2\bigg({\frac{d^2\omega}{dk^2}}\bigg)_{k_0},
\end{equation}
where using the definitions,
\begin{equation}
\label{math:1.5}
\bigg({\frac{d\omega}{dk}}\bigg)_{k_0}=v_g,\quad\quad\frac{1}{2}\bigg({\frac{d^2\omega}{dk^2}}\bigg)_{k_0}=\beta
,\quad\quad k-k_0=k^\prime.
\end{equation}
equation (3) can be written as,
$$f(x,t)=e^{\,ik_0x-i\omega(k_0)t}\int_{-\infty}^\infty dk^\prime
\,e^{-\alpha{k^\prime}^2}e^{\,ik^\prime(x-v_gt)}
\,e^{-i{k^\prime}^2\beta t}$$
\begin{equation}
\label{math:1.6} =e^{\,ik_0x-i\omega(k_0)t}\int_{-\infty}^\infty
dk^\prime \,e^{\,ik^\prime(x-v_gt)}\,e^{-(\alpha+i\beta
t){k^\prime}^2}.
\end{equation}
Now completing the square root in exponent and integration gives,
\begin{equation}
\label{math:1.7}f(x,t)=e^{\,i\big[k_0x-\omega(k_0)t\big]}\,\bigg(\frac{\pi}{\alpha+i\beta
t}\bigg)^{\frac{1}{2}}\,e^{-\big[\frac{(x-v_gt)^2}{4(\alpha+i\beta
t)}\big]}.
\end{equation}
Therefore one find,
\begin{equation}
\label{math:1.8}|f(x,t)|^2=\bigg(\frac{\pi^2}{\alpha^2+\beta^2
t^2}\bigg)^{\frac{1}{2}}e^{-\big[\frac{\alpha(x-v_gt)^2}
 {2(\alpha^2+\beta^2
t^2)}\big]},
\end{equation}
which is the profile of the wave in position space. The quantity
which in $t=0$ was $\alpha$, now has became $\alpha+\frac{\beta^2
t^2}{\alpha}$ and this is the notion of broadening. Therefore,
\begin{equation}
\label{math:1.9}Broadening\propto \bigg(1+\frac{\beta^2
t^2}{\alpha^2}\bigg)^{1\over2}.
\end{equation}
This relation shows that a wave packet with width $(\Delta x)_{0}$
in $t=0$ after propagation will have the following width,
\begin{equation}
\label{math:1.10}(\Delta x)_{t} = (\Delta
x)_{0}\bigg(1+\frac{\beta^2 t^2}{\alpha^2}\bigg)^{1\over2}.
\end{equation}
\subsection{ Wave Packet Propagation in Quantum Gravity} As has been
indicated, when one considers gravitational effects, usual
uncertainty relation of Heisenberg should be replaced by,
\begin{equation}
\label{math:2.1}\Delta x\geq\frac{\hbar}{\Delta
p}+\frac{\alpha^\prime l_p^2\Delta p}{\hbar}.
\end{equation}
It is important to note that there are more generalization which
contain further terms in right hand side of equation (11)( see[14]),
but in some sense equation (11) has more powerful physical grounds.
So as a first step analysis we consider the above simple form of
GUP. Suppose that
$$\Delta x\sim x,\quad\quad\Delta p\sim p,\quad\quad p=\hbar k,\quad\quad x=\bar{\lambda}=\frac{\lambda}{2\pi}.$$
Therefore one can write,
\begin{equation}
\label{math:2.2}\bar{\lambda}=\frac{1}{k}+\alpha^\prime l_p^2\,k
\qquad and \qquad \omega=\frac{c}{\bar{\lambda}}.
\end{equation}
In this situation  the dispersion relation becomes,
\begin{equation}
\label{math:2.3}\omega=\omega(k)=\frac{kc}{1+\alpha^\prime
l_p^2\,k^2}.
\end{equation}
This relation can be described in another viewpoint. By expansion of
$\bigg(1+\alpha^\prime l_p^2\,k^2\bigg)^{-1}$ and neglecting second
and higher order terms of $\alpha^{\prime}$, we find that
$\omega=kc\big(1-\alpha^\prime l_p^2\,k^2\big)$. This can be
considered as $\omega=k^{\prime}c$ where
$k^{\prime}=k\big(1-\alpha^\prime l_p^2\,k^2\big)$. Now one can
define a generalized momentum as $p=\hbar k^{\prime}=\hbar
k\big(1-\alpha^\prime l_p^2\,k^2\big)$. It is possible to consider
this equation as $p=\hbar^{\prime}k$ where
$\hbar^{\prime}=\hbar\big(1-\alpha^\prime l_p^2\,k^2\big)$. So one
can interpret it as a wave number dependent Planck "constant". In
the same manner group velocity becomes,
\begin{equation}
\label{math:2.4}v_g=\frac{d\omega}{dk}\bigg|_{k=k_0}=\frac{c(1-\alpha^\prime
l_p^2\,k^2)}{(1+\alpha^\prime l_p^2\,k^2)^2}\bigg|_{k=k_0}.
\end{equation}
Up to first order in $\alpha^{\prime}$ this relation reduces to
$v_{g}\approx c\big(1-3\alpha^{\prime}l_{p}^{2}k_{0}^{2}\big)$.\\
A little algebra gives $\beta$ as follow
\begin{equation}
\label{math:2.5}\beta=\frac{1}{2}\bigg(\frac{d^2\omega}{dk^2}\bigg)\bigg|_{k=k_0}
=\frac{-3\alpha^\prime l_p^2c\,k(1+\alpha^\prime l_p^2
\,k^2)^2+4{\alpha^\prime}^2l_p^4c\,k^3(1+\alpha^\prime l_p^2
\,k^2)}{(1+\alpha^\prime l_p^2\,k^2)^4}\Bigg|_{k=k_0},
\end{equation}
which up to first order in $\alpha^{\prime}$ reduces to
 $\beta \approx-3\alpha^\prime l_p^2ck_{0}$. It is evident that when
$\alpha^\prime\rightarrow0$ then $\beta\rightarrow0$ and
$v_g\rightarrow c$. The same analysis which has leads us to equation
(10), now gives the following result,
\begin{equation}
\label{math:2.6}(\Delta x)_{t} = (\Delta
x)_{0}\Bigg(1+\frac{1}{\alpha^2}\bigg(\frac{-3\alpha^\prime
l_p^2c\,k_{0}(1+\alpha^\prime l_p^2
\,k_{0}^2)^2+4{\alpha^\prime}^2l_p^4c\,k_{0}^3(1+\alpha^\prime l_p^2
\,k_{0}^2)}{(1+\alpha^\prime l_p^2\,k_{0}^2)^4}\bigg)^2
t^2\Bigg)^{1\over2}.
\end{equation}
If one accepts that $\alpha^{\prime}$ is negative constant (
$\alpha^{\prime}<0$), then group velocity of the wave packet becomes
greater than light velocity. This is evident from equation (14) and
is reasonable from varying speed of light models. In fact if
$|\alpha^{\prime}|k^{2}l^{2}_{p}\ll1$, one recover usual quantum
mechanics but when $|\alpha^{\prime}|k^{2}l^{2}_{p}\approx1$, Planck
scale quantum mechanics will be achieved. Based on this argument,
equation (16) shows that in quantum gravity there exists a more
broadening of wave packet due to gravitational effects. Up to first
order in $\alpha^{\prime}$, this equation becomes,
\begin{equation}
\label{math:2.6}(\Delta x)_{t} = (\Delta
x)_{0}\Bigg(1-\frac{3\alpha^\prime
l_p^2c\,k_{0}t^2}{\alpha^2}\Bigg)^{1\over2}.
\end{equation}
Now using equation (13), one can write the dispersion relation as
the following form also,
\begin{equation}
\label{math:2.7}\omega(p)=\frac{\hbar pc}{\hbar^2+\alpha^\prime
l_p^2\,p^2},
\end{equation}
or
\begin{equation}
\label{math:2.8}E^\prime=\hbar \omega(p)=\frac{p
c}{1+\alpha^\prime\Big(\frac{l_pp}{\hbar}\Big)^2}.
\end{equation}
It is evident that if $\alpha^\prime\longrightarrow0$ Then
$E^\prime\longrightarrow E=pc$  and
$\omega(p)\longrightarrow\omega=\frac{pc}{\hbar}$. These dispersion
relations provide a framework for definition of generalized momentum
operator and generalized Klein-Gordon equation.
\section{ A Generalized Klein-Gordon Equation}
Now consider the following integral which has been used
heuristically by Schr\"{o}dinger to find his equation[1],
\begin{equation}
\label{math:3.1}\psi(x,t)=\frac{1}{\sqrt{2\pi\hbar}}\int{dp\,
\phi(p)e^{i(px-E^\prime t)/\hbar}}.
\end{equation}
Differentiation twice relative to $x$ gives,
\begin{equation}
\label{math:3.2}\frac{\partial^2\psi(x,t)}{\partial
x^2}=-\frac{1}{\hbar^2}\times\frac{1}{\sqrt{2\pi\hbar}}\int{dp\,p^2\,
\phi(p)e^{i(px-E^\prime t)/\hbar}},
\end{equation}
while differentiation relative to $t$ leads to
\begin{equation}
\label{math:3.3}\frac{\partial^2\psi(x,t)}{\partial
t^2}=-\frac{c^2}{\hbar^2}\times\frac{1}{\sqrt{2\pi\hbar}}\int{dp\,\Bigg(\frac{p}{1+\alpha^\prime\Big(\frac{l_pp}
{\hbar}\Big)^2} \Bigg)^2 \phi(p)e^{i(px-E^\prime t)/\hbar}}.
\end{equation}
It is evident that if $\alpha^\prime\longrightarrow0$, then
$\frac{\partial^2\psi(x,t)}{\partial
t^2}=c^2\frac{\partial^2\psi(x,t)}{\partial x^2}$ which is the usual
wave equation. Now expansion of integrand in equations (19) and (20)
gives,
$$\frac{\partial^2\psi(x,t)}{\partial
t^2}=-\frac{c^2}{\hbar^2}\times\frac{1}{\sqrt{2\pi\hbar}}\int{dp\,p^2\bigg(1-\alpha^\prime\Big(\frac{l_pp}
{\hbar}\Big)^2+O({\alpha^\prime}^2)-\cdots \bigg)^2
\phi(p)e^{i(px-E^\prime t)/\hbar}}$$
\begin{equation}
\label{math:3.4}=-\frac{c^2}{\hbar^2}\times\frac{1}{\sqrt{2\pi\hbar}}\int{dp\,\bigg(p^2-\frac{2\alpha^\prime
l_p^2}{\hbar^2}p^4+O({\alpha^\prime}^2)+\cdots \bigg)
\phi(p)e^{i(px-E^\prime t)/\hbar}}.
\end{equation}
The first order terms in $\alpha^{\prime}$, satisfy the following
equation,
\begin{equation}
\label{math:3.5}\frac{\partial^2\psi(x,t)}{\partial
t^2}=c^2\frac{\partial^2\psi(x,t)}{\partial x^2}-2\alpha^\prime
l_p^2c^2\frac{\partial^4\psi(x,t)}{\partial x^4}.
\end{equation}
This is a generalized wave equation and in some sense can be
considered as generalized Klein-Gordon equation for a massless
particle. The solution of this equation gives the correct profile of
the wave in Planck scale. Adler and Santiago in their paper[21] have
indicated that: "It has long been a supposed fact of life that the
differential equations of physics are first or second order. This is
well born out by experiences in that classical mechanics, classical
electromagnetism, general relativity, non-relativistic quantum
mechanics, relativistic quantum mechanics, and all the equations of
the standard model of particles are at most of second order. But it
may well be that we also have developed an unjustified bias in favor
of second order equations due to mathematical convenience. As such
it is particularly interesting to consider higher order equations
with all their inherent dangers and difficulties with boundary
conditions". In our opinion equation (22) is one of the mentioned
higher derivative equations. For a massive particle(or field),
equation (22) can be written as,
\begin{equation}
\label{math:3.6}-\hbar^2\frac{\partial^2\psi(x,t)}{\partial
t^2}=\bigg(-\hbar^2c^2\frac{\partial^2}{\partial
x^2}+2\alpha^\prime\hbar^2 l_p^2c^2\frac{\partial^4}{\partial x^4}+
m_{0}^{2}c^{4}\bigg)\psi(x,t).
\end{equation}
Note that this is only the first order approximation equation since
we have considered only the first order term in (22).\\
In the language of operators, usual Klein-Gordon equation can be
written as,
\begin{equation}
\label{math:3.7}-\hbar^2\frac{\partial^2\psi(x,t)}{\partial
t^2}=\bigg(c^{2}p^{2}_{op}+ m_{0}^{2}c^{4}\bigg)\psi(x,t),
\end{equation}
where $ p_{op} = \frac{\hbar}{i} \frac{\partial}{\partial x}$. Now
the generalized momentum operator in GUP up to first order in
$\beta^{\prime}$, takes the following form,
\begin{equation}
\label{math:3.8}
p^{(GUP)}_{op}=\frac{\hbar}{i}\Bigg(1+\beta^{\prime}
\Big(\frac{\hbar}{i} \frac{\partial}{\partial
x}\Big)^{2}\Bigg)\frac{\partial}{\partial x},
\end{equation}
or
\begin{equation}
\label{math:3.9}
\bigg(p^{(GUP)}_{op}\bigg)^{2}=-\hbar^2\frac{\partial^2}{\partial
x^2}+2\beta^\prime\hbar^4\frac{\partial^4}{\partial x^4}
\end{equation}
and this operator immediately gives (25) with
$\beta^{\prime}=\alpha^{\prime}\frac{l^{2}_{p}}{\hbar^2}$ . The
generalization to 3-dimensional case is straightforward. It is
important to note that one can consider from the beginning a
generalized form of momentum operator. In this case equation
$p=\hbar k$ should be replaced by $p=\hbar k^\prime$ where
$k^{\prime}=k\big(1-\alpha^\prime l_p^2\,k^2\big)$. In this
situation equation (20) should be modified by this generalized
momentum and we find ordinary Klein-Gordon equation but now with
generalized operators.
\section{Summary and Conclusions}
In this paper we have
shown that:
\begin{enumerate}
\item[1-] In quantum gravity, wave packet broadening is more than
corresponding broadening of wave packet in usual quantum mechanics
because of inherent gravitational uncertainties.
\item[2-] There are generalized dispersion relations in quantum
gravity which lead to generalized group velocity, generalized
momentum operator and generalized Klein-Gordon equation.
\item[3-] One can describe the generalized dispersion relations
being as a consequence of generalized de Broglie principle or
varying Planck constant( as a function of momentum ).
\item[4-] Our analysis shows that gravitational uncertainty
principle leads to varying speed of light scenario in such a way
that speed of light in early universe was greater than its present
value.
\item[5-]It seems that in Planck scale quantum mechanics, higher
order derivatives will be appear in equations. This is seen in
generalized Klein-Gordon equation and is a result of generalized
momentum operator.
\item[6-] Our approach provides foundations for constructing a
gravitational quantum mechanics. This is the subject of our
forthcoming work.
\end{enumerate}
{\bf Acknowledgment}\\
This Work has been supported partially by Research Institute for
Astronomy and Astrophysics of Maragha, Iran.

\end{document}